\documentclass[runningheads]{llncs}
\usepackage{graphicx}
\usepackage{cite}
\usepackage{wrapfig}
\usepackage{paralist}
\usepackage{booktabs}
\usepackage{color}
\usepackage{ifthen}
\newboolean{showcomments}
\setboolean{showcomments}{true} 
\ifthenelse{\boolean{showcomments}}
    {
        \newcommand{\per}[1]{\textcolor{blue}{{\it [Per says: #1]}}}
        \newcommand{\johan}[1]{\textcolor{cyan}{{\it [Johan says: #1]}}}
    }
    {
        \newcommand{\per}[1]{}
        \newcommand{\johan}[1]{}
    }

\begin{document}
%
\title{Collaboration in Open Government Data Ecosystems: Open Cross-sector Sharing and Co-development of Data and Software}
%
\titlerunning{Collaboration in Open Government Data Ecosystems}
%
\author{Johan Lin{\aa}ker\inst{1} \and Per Runeson\inst{1}}
%
%
\institute{Lund University, Ole Römers väg 3, Lund, Sweden
\email{\{johan.linaker,per.runeson\}@cs.lth.se}\\}
\maketitle              
\begin{abstract}
\textit{Background}: Open innovation highlights the potential benefits of external collaboration and knowledge-sharing, often exemplified through Open Source Software (OSS). The public sector has thus far mainly focused on the sharing of Open Government Data (OGD), often with a supply-driven approach with limited feedback-loops. We hypothesize that public sector organizations can extend the open innovation benefits by also creating platforms, where OGD, related OSS, and open standards are collaboratively developed and shared.
\textit{Objective}: The objective of this study is to explore how public sector organizations in the role of platform providers facilitate such collaboration in the form of OGD ecosystems and how the ecosystem's governance may be structured to support the collaboration.
\textit{Method}: We conduct an exploratory multiple-case study of two such ecosystems, focused on OGD related to the Swedish labor market and public transport sector, respectively. Data is gathered through interviews, document studies, and prolonged engagement at one of the platform providers.
\textit{Results}: The study presents governance structure and collaboration practices of the two ecosystems and discusses how these contribute to the platform providers' goals.
The case studies highlight the need for platform providers to take an active and multi-functional role in enabling the sharing of data and software from and between the members of the ecosystem.
\textit{Conclusions}: We conclude that OGD ecosystems offer public sector organizations a possibility to catalyze the potential innovation output of OGD, but that it requires investment and adoption of an open and collaborative mindset.

\keywords{Open Government Data  \and Open Source Software \and Open Standard \and Ecosystem \and Public Sector.}


\end{abstract}

%
%
\section{Introduction}
Open Innovation has been widely adopted among software companies through the use of Open Source Software (OSS) as a means to share costs and accelerate innovation~\cite{munir2015open}. In the public sector, however, the focus has been more on sharing of Open Government Data (OGD) as a way to drive innovation~\cite{attard2015systematic} and less on OSS\cite{borg2018digitalization}.
To catalyze the potential innovation output~\cite{attard2015systematic}, data providers and data users may form a business ecosystem around the OGD~\cite{zuiderwijk2014innovation}. Actors within an OGD ecosystem together create a value network, where the OGD is enriched from raw data into valuable content to be used in new products and services~\cite{lindman2015business}. Interaction is usually limited to feedback on the quality and what data-sets to release next~\cite{m2017open, immonen2014requirements, dawes2016planning}, although the need for more collaboration is highlighted~\cite{sieber2015civic, rudmark2019harnessing}.

Looking at Open Source Software (OSS) ecosystems (also commonly referred to as communities)~\cite{franco2017open}, actors collaborate to a much a higher degree through the co-development of the OSS. Together they evolve the OSS in an open setting where new functionality is continuously asserted, discussed, and implemented and thereby they accelerate the innovation and development beyond what any single member of the ecosystem could perform alone~\cite{munir2015open}.

We hypothesize that by adopting the collaborative practices, OGD ecosystems would be able to elicit similar benefits~\cite{RunesonNIER2019, sieber2015civic}. Further, we hypothesize that both the sharing and adoption of OGD would be catalyzed by extending collaboration to include the development of related standards, APIs, and supporting tools, frameworks, and example applications as OSS~\cite{zuiderwijk2014innovation, immonen2014requirements, rudmark2019harnessing}. With these characteristics, we define an OGD ecosystem as \textit{a networked community of organizations, which base their relations to each other on a common interest in an underpinning technological platform consisting of OGD and related OSS and open standards, and collaborate through the exchange of information, resources and artifacts}, adapted from~\cite{zuiderwijk2014innovation, jansen2020focus}.

Existing research has mainly focused on the collaborative practices used in OSS ecosystems~\cite{alves2017software, jansen2020focus}, and has in terms of OGD ecosystems~\cite{attard2015systematic} been limited, both regarding collaboration on OGD or any related software or standard, even though identified as a need~\cite{sieber2015civic, oliveira2019investigations, rudmark2019harnessing}. The \textbf{research goal} of this study is to therefore to \textit{explore how collaboration in OGD ecosystems may be facilitated from the platform providers' point of view and how the ecosystems' governance may be structured to support the collaboration}. We find this as an interesting perspective as trust towards the platform provider is pivotal in order to enable collaboration and growth of an ecosystem~\cite{lindman2015business, jansen2014measuring}. 

This paper presents a multiple-case study~\cite{runeson2012casestudy} of two OGD ecosystems in which OSS, open standards, and related collaborative practices are adopted, aiming to foster collaboration and increase the adoption of OGD. The two ecosystems are initiated and governed by public sector organizations in the role of platform providers and focus on OGD related to the Swedish labor market and public transport sector. We present the governance structure and collaboration practices of these ecosystems and discuss how these contribute to the platform providers' goals.

\section{Background}
Below we provide an overview of OGD and software (including OSS) ecosystems and present a governance model used 
in the analysis of the two case studies.

\subsection{Ecosystems for Open Source Software and Government Data}
Software ecosystems is a rather mature\cite{alves2017software, jansen2020focus} but yet expanding field of research. Originating from the field of business ecosystems (and in turn its biological ancestor)~\cite{iansiti2004keystone}, it offers a lens for analyzing how networked communities of organizations collaborate around their common interest in a central software technology~\cite{mhamdia2013performance}. Other definitions refer to technological platforms underpinning the ecosystem~\cite{jansen2020focus}, which in a wider sense may consist of either technologies, products, or services, serving in the common interest of the ecosystem.

For OSS ecosystems, the OSS project makes up the technological platform underpinning its community which we refer to as the ecosystem~\cite{franco2017open}. Similarly, OGD and its related APIs can also be viewed as a platform, underpinning the surrounding ecosystem of actors~\cite{zuiderwijk2014innovation}. OGD ecosystems and similar concepts~\cite{SushaTGPPP17, sieber2015civic} are however not as well explored as software and OSS ecosystems. 



Oliveira et al.~\cite{oliveira2019investigations} define a data ecosystem as \textit{``socio-technical complex network in which actors interact and collaborate with each other to find, archive, publish, consume, or reuse data as well as to foster innovation, create value, and support new businesses}. OGD ecosystems, or Government Data Ecosystems as referred to by Oliveira et al.~\cite{oliveira2019investigations}, are based on OGD initiatives and focused on promoting the use and publication of OGD. 

In terms of roles in these ecosystems, a general distinction can be made between data providers and data users~\cite{zuiderwijk2014innovation}. The roles can be further refined into data providers, service providers, data brokers, application developers, application users, and infrastructure and tool providers~\cite{immonen2014requirements, kitsios2017business}. The data provider is usually constituted by a public-sector organization~\cite{oliveira2019investigations}. Services or functions needed include an infrastructure to share the data (preferably from multiple providers), documentation, tools for application developers, help in finding use-cases, as well as the possibility to discuss, provide feedback and make requests~\cite{immonen2014requirements, zuiderwijk2014innovation, dawes2016planning}. A general observation regards the need for improved feedback-loops, collaboration and a more demand-driven publication of OGD~\cite{rudmark2019harnessing, zuiderwijk2014innovation, dawes2016planning, m2017open}. 

This study aims to investigate how such collaboration may be facilitated, both in terms of sharing and co-developing OGD, related OSS, and open standards. We hypothesize that this will provide further opportunities and benefits of what open innovation has to offer to the platform providers and their ecosystems~\cite{munir2015open, RunesonNIER2019}.



\subsection{Ecosystem Roles and Governance Structure}
Existing research is limited in regard to governance in OGD ecosystems~\cite{oliveira2019investigations}. Governance has received more attention for software ecosystems~\cite{alves2017software}. Three types of roles are commonly referred to~\cite{jansen2020focus, iansiti2004keystone}. The first role is that of the \textit{platform provider} who is the owner and supplier of the platform and thereby also usually the orchestrator of the ecosystem. As an orchestrator, the platform provider also decides on the governance model for the ecosystem, meaning the ways in which it maintains control and decides on the direction, but also on the governance structure, meaning ``the distribution of rights and responsibilities among the [ecosystem's members], and the rules and protocols that need to be followed in order to make decisions regarding the [ecosystem]''~\cite{alves2017software}. 

\textit{Keystone} and \textit{Niche players} are two other roles within an ecosystem. A keystone is an actor who nurtures a symbiotic relationship with the ecosystem and its other actors, looking to actively improve its health~\cite{jansen2014measuring}. Usually, they have a close connection with the platform provider, who also may be referred to as a keystone if it has similar symbiotic intents. Niche players are actors focused more on a specific niche of the market, or use-case, and is primarily a user of the resources provided by the ecosystem~\cite{iansiti2004keystone}.

For OSS ecosystems, the platform provider can be the owner of the OSS project, usually either a software vendor or the ecosystem of actors directly or via a proxy organization (e.g., a foundation)~\cite{riehle2012single}. Governance, however, does not have to be aligned with the ownership. In more autocratic ecosystems, it can be centered around a vendor or individual, while more democratic ecosystems it is distributed~\cite{de2013evolution}. In the latter case, control of the OSS project is usually maintained by a central group of actors who have gained a level of influence by proving merit, building trust, and social capital through contributions to the OSS project.

\subsection{Governance Model for Open Government Data Ecosystems}\label{sec:rw:model}

\begin{figure}[t]
\begin{center}
\includegraphics[scale=0.3]{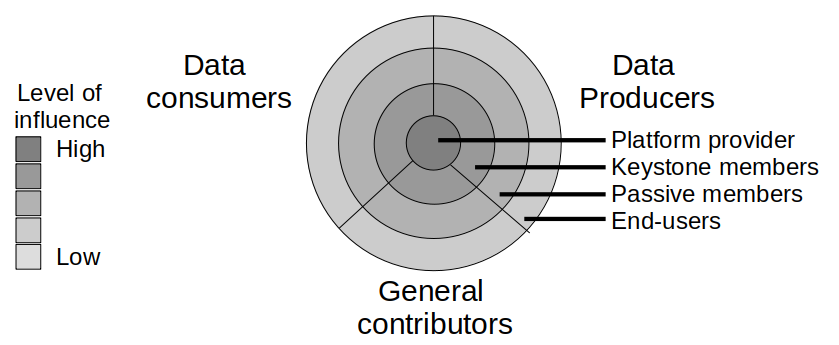}
\vspace{-0.5cm}
\caption{Overview of the proposed governance model adapted from Nakakoji et al.~\cite{nakakoji2002evolution}.}
\label{fig:GovModel}
\vspace{-0.8cm}
\end{center}
\end{figure}

A popular way of illustrating the governance structure of an OSS ecosystem is the Onion model~\cite{nakakoji2002evolution}, where the center is those in control (see Fig.~\ref{fig:GovModel}). The closest layers may be those who contribute actively to the project and thereby maintain an influence although not in direct control. For each outer layer, actors become less active in terms of contributions and thereby decrease in influence on the OSS project. Robles et al.~\cite{robles2019setting} recently applied the model in a case study on the X-Road OSS project, an originally Estonian eGovernment project for creating a data-sharing infrastructure, which now is governed jointly by Estonian and Finnish government agencies. The project is centrally controlled, and contributions are primarily made by companies on behalf of, and paid by, the government agencies. 

For OGD ecosystems, we consider the core to be occupied by the \textit{platform provider} (see Fig.~\ref{fig:GovModel}), which is either the government entity (or entities in collaboration) which provide OGD via a software platform where APIs and supporting tools, frameworks, and example applications are available as OSS. Depending on the specific ecosystem structure a number of layers follow. In layers closest to the core are the \textit{Keystone members} including actors that are of special importance to the platform provider and the overall health of the ecosystem~\cite{jansen2014measuring}. In the following layer, \textit{Passive members} of similar roles may be found although these are more focused on addressing their specific niche or use-case. In the last and outer layer are the \textit{End-users} of the OGD, either directly or via the proxy of applications and services produced by the actors in the inner layers of the model.

As illustrated in Fig.~\ref{fig:GovModel}, actors can in general also be divided between three groups in terms of their usage of and contribution to the platform:
\begin{compactitem}
  \item A \textit{Data consumer} uses the data available via the platform. 
  \item A \textit{Data producer} contributes data, either actively via a donation to the platform, or passively by letting the platform provider collect data from them, which is then made available via the platform.
  \item A \textit{General contributor} is not necessarily a consumer or producer of data, but in some other way contributes to the platform and health of the ecosystem, e.g., through knowledge sharing or contributing new or to existing OSS projects related to the platform.
\end{compactitem}

Actors can be further categorized based on their type of operations, including e.g., service providers, application developers, data brokers, infrastructure and tool providers, and potentially additional data providers~\cite{immonen2014requirements, lindman2015business}. We expect this type of categorization to be dependent on each ecosystem.


\section{Research Design}
An exploratory multiple-case study~\cite{runeson2012casestudy} was conducted to investigate two instances of OGD ecosystems. Case1 is the JobTech Dev ecosystem, initiated and facilitated by The Swedish Public Employment Service. Case2 is the Trafiklab ecosystem, initiated and facilitated by Samtrafiken. The unit of analysis is the ecosystems' governance~\cite{alves2017software}.

We use an adapted version of the Onion model (see section~\ref{sec:rw:model}) to structure and analyze the findings of the two ecosystems' governance. The model is commonly used for describing the governance in OSS ecosystems~\cite{nakakoji2002evolution} and was recently applied to a government-initiated OSS ecosystem~\cite{robles2019setting}.

The research effort was initiated with Case 1, where the first author of this study is embedded as an action researcher, as a part of a long-term research project. The researcher was hence able to generate in-depth knowledge through prolonged engagement along with access to extensive documentation. The documentation along with field notes could be used to triangulate findings along with three semi-structured interviews. To ensure construct validity~\cite{runeson2012casestudy}, we based the questionnaire on earlier work on assessing the governance structure of software ecosystems~\cite{alves2017software, jansen2020focus}. The interviewees were the platform's product owner, community manager, and policy strategist.

Data gathering from Case 1 was performed before any intervention had been introduced from the action research. To avoid researcher bias, peer-debriefing between the first and second authors was performed~\cite{runeson2012casestudy}.

For Case 2, data were gathered in a similar manner through a semi-structured interview with the platform's product manager, using the same questionnaire. All interviews were audio-recorded with additional notes taken. A threat regarding the reliability concerns that only the first author conducted the interviews~\cite{runeson2012casestudy}. To mitigate the threat, member-checking was performed in both cases where synthesized findings were presented to all interviewees who were asked for correctness, misinterpretations, and redundancy. 

\section{Results}\label{sec:results}

Below we present the results from our two studied cases, JobTech Dev and Trafiklab.

\begin{figure}[t]
\begin{center}
\includegraphics[width=\columnwidth]{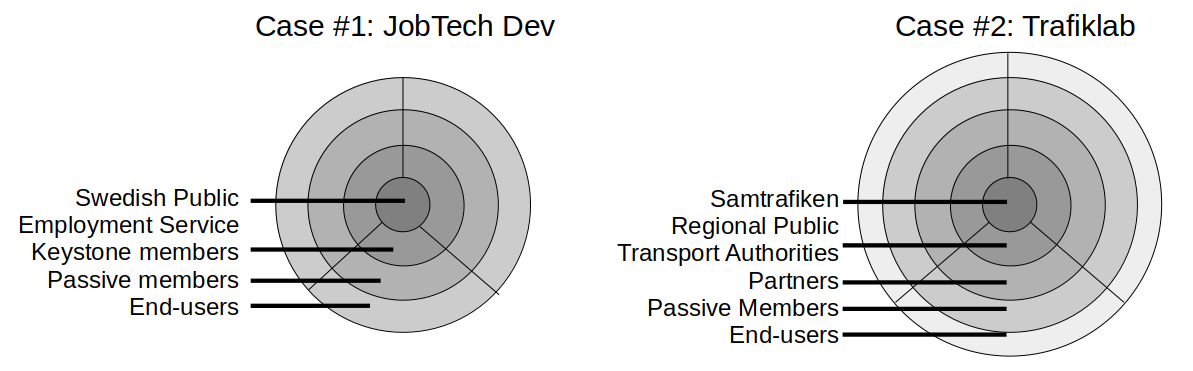}
\label{fig:Cases}
\vspace{-0.5cm}
\caption{Overview of the governance structure of the two cases in this study, JobTech Dev (left) and Trafiklab (right), based on the governance model presented in section~\ref{sec:rw:model}.}
\vspace{-0.8cm}
\end{center}
\end{figure}

\subsection{Case 1: JobTech Dev}
JobTech Dev is an ecosystem, initiated in 2018, bringing actors together, operating within or in relation to the Swedish labor market to collaborate on a common platform of OGD, connected APIs, and complementary OSS projects. The ecosystem and its platform are developed and orchestrated by the Swedish Public Employment Service (SPES), a Swedish national government agency responsible for enabling the match-making between job-seekers and employers on the labor market. JobTech Dev was created as a part of this mission with the intention to enable the actors in the ecosystem to accelerate their innovation process, improve their services, and thereby improving the digital match-making on the labor market.

\subsubsection{Platform Content}
The platform consists of four parts: \textit{Jobs, Taxonomy, Career}, and \textit{Search}.

Jobs is a service intended to collect all available job advertisements on the Swedish labor market and make these available through an API. The online ads are collected from the ten organizations in the labor market, providing the majority of the available advertisements.

Taxonomy is a collection of skills and job titles and relationships between them. The data set is developed and maintained within SPES. By opening up the data through APIs, the actors in the ecosystem are enabled to ``speak the same language'' enabling e.g., improved reporting and statistics and match-making between job advertisements and job-seekers' resumes.

Career is (unlike Taxonomy and Jobs) not OGD. It is a service where job-seekers can store their resumes on a central location in an encrypted format. The job-seekers can grant and withdraw permission to organizations, e.g., recruitment firms, social networks, and insurance firms to access their information. The service is based on the MyData principles\footnote{https://mydata.org/guiding-principles/} and enables job-seekers to only have to maintain one copy of their resume and to distribute and manage their data with kept control over their integrity and privacy.

Search is an OSS search engine that enables actors to search among available job advertisements. Search is available both through an API and as an OSS project which can be adopted and integrated by the ecosystem's actors.

\subsubsection{Ecosystem Governance Structure}
In terms of operations, the members of the ecosystem can generally be categorized within one of the areas: 1) recruiting and staffing firms, 2)~education and guidance providers, 3)~national, regional or local governments, 4)~workers' unions, 5)~employers' associations, 6)~job advertisers, and 7)~job seekers. Depending on the category, a member's interest in the platform may be limited to certain parts of the platform.

Considering the governance structure, SPES is positioned in the center as the platform provider orchestrating and governing the ecosystem (see Fig.~\ref{fig:Cases}, left). SPES ultimately decides on requirements and road-map for the platform, including what data to make available, when, and how. They perform the necessary development and maintenance and provide the infrastructure needed to access and use the data. 

Outside of SPES are the members whose opinions may be considered as extra important for SPES in terms of developing the platform and growing the ecosystem. These members may, e.g., have a large user base, or valuable competencies and resources, and thereby contribute to the health of the ecosystem. In the second layer are the general members and in layer three the end-users. Each layer is viewed to potentially consist of members from all types of operations. 


\subsubsection{Orchestration and Collaboration}
Due to the limited internal resources, SPES does not have the capacity to maintain formal and direct relationships with all ecosystem members.
Teams carrying out development inside SPES therefore primarily work and communicate through close relationships with the key members to optimize the impact. However, SPES is striving to adopt an open development model and maintain an open dialogue where the whole ecosystem (including all layers in Fig~\ref{fig:GovModel}) can influence the direction of the platform.

Anyone can, for example, request and discuss a new feature, an API, or data-set through a synchronous open communication platform or by attending occasional meetups arranged by SPES. It is also possible to contribute to the development, as all source code for the APIs is available as OSS. To lower adoption barriers of the data provided via the platform, example applications are developed and released as OSS. Members consuming the data, specifically startups, have expressed the value provided by these examples as it helps them understand use-cases and accelerate their development. Contributions to the OSS have been limited to bug reports and feature requests, while the intention is to encourage and enable members to contribute both new projects and to exist.

In terms of data, there are examples -- although limited -- of members producing and contributing directly to the platform. One example is a set of soft skills and their relationship to different job titles, which was contributed to the Taxonomy part of the platform. Processes are not yet established for how these types of contributions should be managed; a process more simpler if considering OSS. In this case, a formal contract was established between the two parties. 

Close collaborations and direct dialogues with key members have been important to establish the ecosystem and gain general acceptance. For example, the collection of all job-ads on the market and making these freely available, made incumbents offering recruiting and staffing services initially question the intent from SPES as well as the suggested benefits. SPES views the commoditization of job-advertisement data, as with the ecosystem at large, as a way to push the actors working with digital match-making and guidance services forward, nurturing innovation and lowering barriers to entry for new actors.

After a more than two-year process, even the more conservative incumbents started to accept the ecosystem and see potential benefits with it. A formal collaboration was initiated between SPES and the ten largest job advertisement providers where they agreed to allow job-advertisements to be collected. Once collected, the ads are converted to an industry-specific open standard, and then enriched with metadata such as date of publishing and deadline for applications. A compromise was reached to only provide a ``stub'' of the advertisements, meaning that only that job title, metadata, and a link to the original advertisement would be provided through the platform's API.

\subsection{Case 2: Trafiklab}
Trafiklab is an ecosystem, initiated in 2011, that brings actors within the Swedish public transport sector together to collaborate on a platform with open traffic data, connected APIs, and complementary OSS projects. The ecosystem's vision is to facilitate the creation of new services that makes it easier and more attractive to travel with public transport. The ecosystem and its platform are developed and orchestrated by Samtrafiken, a corporate entity co-owned by all the regional public transport authorities and most of the commercial transport operators in Sweden. The commercial transport operators also have the option of being a partner to Samtrafiken.

\subsubsection{Platform Content}
The platform consists of data-sets and APIs, either maintained by Samtrafiken or independently by members of the ecosystem. All data hosted on the Trafiklab-platform is released with a custom license based on the principles of the Creative Commons Attribution-license.

Four APIs provide static and real-time data on public transport, related to, for example, time-tables and interruptions. This data is currently made available in two types of standard formats, maintained by Samtrafiken and gathered from the regional public transport authorities and private operators in accordance with a government directive. Two further APIs provide time-table data for a trip-planner, an externally procured product that is offered for free to the ecosystem.

Certain APIs are maintained by other organizations, both public and private, and made available on the Trafiklab-platform. Data includes time-table and service data from Stockholm Public Transport and traffic information from the Swedish Transport Administration. The platform also links to related APIs that are maintained and hosted by other organizations. These include data from regional public transport authorities, local counties, and private entities.

\subsubsection{Ecosystem Governance Structure}
In terms of operations, the members of the ecosystem can generally be categorized within one of the areas: 1)~regional public transport authorities, 2)~private and publicly owned train operators,~3) national, regional, and local governments, 4)~private bus operators, and 5)~private product and service providers. Future plans include integration with related actors, such as taxi operators and rental-service providers of e.g., cars and bikes.

In terms of the governance structure, Samtrafiken is positioned in the center as the platform provider orchestrating and governing the ecosystem (see Fig.~\ref{fig:Cases}, right). Outside of Samtrafiken in the first layer are regional public transport authorities. As these are formal owners of the platform, they have a strong influence on the direction of Trafiklab. In the second layer are the formal partners to Samtrafiken which may include actors with different types of operation. The third layer primarily consists of private product and service providers, while end-users are positioned in the fourth layer.


\subsubsection{Orchestration and Collaboration}
Close relationships are maintained to regional public transport authorities and partners as these are the primary data producers but also consumers. The ecosystem at large has the possibility to report bugs, ask for help, and request and discuss new features, APIs, or data-sets through an asynchronous, open communication platform. Physical meetings can also serve a similar purpose as Samtrafiken frequently hosts hackathons and meetups related to Trafiklab. 

As with SPES, Samtrafiken is transitioning to a more open and collaborative way of engaging with its ecosystem, a need identified in earlier research~\cite{rudmark2019harnessing}. As an example, they are discussing a more formal approach where users can request and vote on what data sets should be prioritized. In regards to OSS, they currently have a number of software development kits and example applications available. Their intention is to develop a new OSS trip planner and share their APIs as OSS along with their internal road-maps for the different parts of the platform. Contributions have been limited to bug reports and feature requests.

Regarding the data, all of the provided data sets originate from data producers within the ecosystem. Depending on the case, Samtrafiken may transform the data to certain standard formats, develop and maintain the necessary APIs, and provide the necessary infrastructure for data consumers. 
A challenge with growing the ecosystem and gaining new data producers is related to standard formats of the data. For smaller actors, it is an expensive process to transform the data, and for Samtrafiken a recognized risk is that data may be destroyed when transformed between standards. Samtrafiken is, therefore, developing an input-portal to enable further actors to share their data on Trafiklab and to automate the transformation process. The portal is specifically intended for actors in areas related to public transport, such as taxi operators and rental-service providers.

The input-portal is a result of a long-term investigation conducted by Samtrafiken and its partners into the future potential and needs for public transport-related OGD. The investigation also rendered in a plan to introduce 12 new data sets by 2021. 
Other than helping data producers to transform their data into different standards, Samtrafiken and its partners within the Trafiklab ecosystem also collaborate on the development of new standards when needed. A standard for tickets and payment transactions was developed in response to difficulties with different proprietary solutions not being able to interface with each other.

\section{Discussion}


\begin{table*}[t!] 
\vspace{-0.3cm}
\centering
\caption{General characteristics of the two ecosystems investigated, JobTech Dev and Trafiklab, in terms of their platform provider and keystones, but also in regards to the data and software provided on their underpinning platforms.}
\label{tbl:ecosystemCharacteristic}
\begin{tabular}{p{2cm} p{5cm} p{5cm}}
\toprule
& \textbf{JobTech Dev} & \textbf{Trafiklab} \\ \midrule

\textbf{Platform provider} & 
A single government agency & 
A public entity co-owned by multiple public sector organizations \\\midrule

\textbf{Keystones} &
Organizations with large user bases, or valuable assets (e.g., data) & 
Owners and partners \\\midrule

\textbf{Data sources} & 
Produced internally \newline
Collected externally \newline 
Contributed from third party & 
Collected externally \newline
Contributed by third party \\\midrule

\textbf{OSS} &
APIs, example applications &
Example applications, toolkits, libraries, frameworks \\
\bottomrule
\end{tabular}
\vspace{-0.3cm}
\end{table*}

JobTech Dev and Trafiklab present both similar and differentiating attributes as OGD ecosystems (see table~\ref{tbl:ecosystemCharacteristic}). Considering the type of governance structure, we observe that JobTech Dev is governed by SPES, a single government agency, while Trafiklab is governed by Samtrafiken, an organization co-owned by all the Swedish public transport authorities, whom all are situated in the inner layer of the governance structure (cf.~\cite{de2013evolution}). SPES has the advantage that it potentially can move quickly but also has the risk that its directive can easily change due to change in the national government. The latter could be a concern for existing and new members whether they can trust the direction and stability of the platform and ecosystem, and thereby if they should invest in platform integrations. 

Samtrafiken, on the other hand, is an investment by several authorities and provides a somewhat neutral body with which commercial traffic operators can become co-owners or create formal partnerships with. This may be a way to ensure trust in the platform provider's commitment and a guarantee for the long-term stability of the ecosystem. Similarities may be drawn to the role of foundations as a proxy-organization and neutral home for OSS projects where actors can collaborate and invest together in a way that benefits them all and with clear charters stating how the project will be technically governed~\cite{de2013evolution}.

X-roads, a governmental OSS project~\cite{robles2019setting}, has a centralized governance model, which can be used for reference. It is clear that they are in control of the ecosystem but at the price of limited contributions from other members. The central governance actors paid the members for their contributions to evolve the OSS. Clearly, there is a trade-off between control and inclination to contribute.


In regards to data sharing, there is also some level of distinction between the two ecosystems. In JobTech Dev, taxonomy data originate from and is provided by SPES, job-advertisement originates from a number of organizations and is collected by SPES, while resume-data is provided on an individual level by the job-seekers. In Trafiklab, all data originate from a third party. Samtrafiken collects, (in some cases transforms) and provides the data on the Trafiklab-platform. The platform's roadmap includes the development of an input-portal to enable further third-parties to contribute their data. 
Both SPES and Samtrafiken thereby show an active role in enabling and incentivizing the members of the respective ecosystems to share their data, either by collecting the data or by enabling a self-service function where the member can provide the data themselves.

Concerning OSS both SPES and Samtrafiken see the value in developing complementary and supporting software as OSS, even though outside contributions thus far have been limited for both. This may, however, be subject to change as both are striving towards adopting a more open and collaborative way of working and collaborating with their ecosystems (cf.~\cite{sieber2015civic, rudmark2019harnessing}). Observations of the X-roads project may generalize to these contexts as well~\cite{robles2019setting}.

Looking at the responsibilities taken on by the platform providers in the investigated ecosystems, they cover many of the roles as reported in literature~\cite{immonen2014requirements, lindman2015business}. Besides being \textit{data providers}, both SPES and Samtrafiken may be described as \textit{data brokers} as they gather, promote, and distribute data from third-party~\cite{immonen2014requirements}, but also data transformers as the transform data between different standards based on ecosystem needs~\cite{lindman2015business}. Another important role is that of \textit{tool providers} as they both develop supporting tools, frameworks, and example applications for their ecosystems~\cite{immonen2014requirements}. Hence, for the collaboration in OGD ecosystems to function the platform provider is required to perform multiple functions compared to traditional OGD ecosystems where tasks are more divided among the actors~\cite{immonen2014requirements, lindman2015business}. In contrast, when entering an existing ecosystem, an organization may consider taking a peripheral less complex role~\cite{rudmark2019harnessing}. This is, however, a trade-off between influence on the platform development and value capture, i.e., if the organization's goals can still be achieved.

As Samtrafiken started their ecosystem in 2011, and SPES in 2018, the former is more mature in its role as a platform provider. For example, it has an existing and active ecosystem with several successful applications and use cases, mechanisms for collecting and enabling actors to share their data, and also practices for enabling and collaboratively designing open standards for the data. SPES has been able to catch up rather rapidly however in terms of platform content due to the dedicated resources, and the existence of both data and software that could be opened up.



\section{Conclusions}
This paper explores the collaboration in OGD ecosystems between actors in terms of sharing and co-development of both OGD, related OSS, and open standards. We focus on the platform provider's point of view and explore how these ecosystems may function in terms of collaboration and governance.

The case studies contrast how the governance structure may differ when the platform provider represents a single or multiple public sector organizations. The case studies further highlight the need for the platform provider to take on an active and multi-functional role in enabling the sharing of data and software from and between the members of the ecosystem. Responsibilities include the creation of processes and technical infrastructure, facilitating standard development and enabling automated transformation between the formats, and co-developing APIs, complementary tools, frameworks, and example applications as OSS.

We hypothesize that this type of collaboration can extend the potential benefits for the platform provider and the ecosystem, including reduced cost and accelerated innovation, but also increased adoption and sharing of data. As this study is limited to exploring the collaboration in two instances of OGD ecosystems, further research is required to validate these hypotheses, create a deeper understanding, and improve the external validity~\cite{runeson2012casestudy}. Readers should consider the context of the platform providers and their ecosystems as reported and adopt an analytical generalization to cases with similar characteristics~\cite{runeson2012casestudy}. 


Several avenues for future research exist. From a platform provider's perspective, for example, \textit{i)} what challenges they experience when ``opening up'' the development of the platform towards the surrounding ecosystem, and \textit{ii)} how the governance and processes for collaborative development should be designed accordingly. From the ecosystem's perspective, including both keystones and niche players, \textit{iii)} what motivates their usage of the platform, sharing of resources and knowledge, as well as active collaboration on the platform's development, \textit{iv)} what challenges do they perceive in these regards, and \textit{v)} how the collaborative development and governance should be designed to best support their participation. Further, \textit{vi)} how do these aspects, as well as dynamics and inter-relationships between the different roles, vary with the up-scaling of the ecosystems, and also between contexts, as with the two cases in this study.

%
%
%
\bibliographystyle{splncs04}
\bibliography{references}
%
\end{document}